\documentclass[prc,aps,showpacs,superscriptaddress,amsmath,amssymb,floatfix,nofootinbib,superscriptaddress]{revtex4}
\usepackage{graphicx}
\usepackage{epsfig}
\newcommand{\fpi}{\mbox{$F_\pi$}}
\newcommand{\qsq}{\mbox{$Q^2$}}
\newcommand{\sigl}{\mbox{$\sigma_{\mathrm{L}}$}}
\newcommand{\sigt}{\mbox{$\sigma_{\mathrm{T}}$}}
\newcommand{\siglt}{\mbox{$\sigma_{\mathrm{LT}}$}}
\newcommand{\sigtt}{\mbox{$\sigma_{\mathrm{TT}}$}}
\newcommand{\eps}{\mbox{$\epsilon$}}
\newcommand{\Lpi}{\mbox{$\Lambda^2_\pi$}}
\newcommand{\Lrho}{\mbox{$\Lambda^2_\rho$}}
\newcommand{\gevsq}{\mbox{GeV$^2$}}

\begin{document}
\verb| |\\  % Seems to fix some latex bug
%\special{papersize=8.5in,11in}
\title{Determination of the pion charge form factor for \qsq=0.60-1.60 \gevsq}

\author{V. Tadevosyan}
\affiliation{Yerevan Physics Institute, 375036 Yerevan, Armenia}
\author{H.P. Blok}
\affiliation{Faculteit Natuur- en Sterrenkunde, Vrije Universiteit, NL-1081 HV 
        Amsterdam, The Netherlands}
\affiliation{NIKHEF, Postbus 41882, NL-1009 DB Amsterdam, The Netherlands}
\author{G.M. Huber}
\affiliation{University of Regina, Regina, Saskatchewan S4S-0A2, Canada}
\author{D. Abbott}
\affiliation{Physics Division, TJNAF, Newport News, Virginia 23606}
\author{H. Anklin}
\affiliation{Florida International University, Miami, Florida 33119}
\affiliation{Physics Division, TJNAF, Newport News, Virginia 23606}
\author{C. Armstrong}
\affiliation{College of William and Mary, Williamsburg, Virginia 23187}
\author{J. Arrington}
\affiliation{Argonne National Laboratory, Argonne, Illinois 60439}
\author{K. Assamagan}
\affiliation{Hampton University, Hampton, Virginia 23668}
\author{S. Avery}
\affiliation{Hampton University, Hampton, Virginia 23668}
\author{O.K. Baker}
\affiliation{Hampton University, Hampton, Virginia 23668}
\affiliation{Physics Division, TJNAF, Newport News, Virginia 23606}
\author{C. Bochna}
\affiliation{University of Illinois, Champaign, Illinois 61801}
\author{E.J. Brash}
\affiliation{University of Regina, Regina, Saskatchewan S4S-0A2, Canada}
\author{H. Breuer}
\affiliation{University of Maryland, College Park, Maryland 20742}
\author{N. Chant}
\affiliation{University of Maryland, College Park, Maryland 20742}
\author{J. Dunne}
\affiliation{Physics Division, TJNAF, Newport News, Virginia 23606}
\author{T. Eden}
\affiliation{Norfolk State University, Norfolk, Virginia 23504}
\affiliation{Physics Division, TJNAF, Newport News, Virginia 23606}
\author{R. Ent}
\affiliation{Physics Division, TJNAF, Newport News, Virginia 23606}
\author{D. Gaskell}
\affiliation{Oregon State University, Corvallis, Oregon 97331}
\author{R. Gilman}
\affiliation{Rutgers University, Piscataway, New Jersey 08855}
\affiliation{Physics Division, TJNAF, Newport News, Virginia 23606}
\author{K. Gustafsson}
\affiliation{University of Maryland, College Park, Maryland 20742}
\author{W. Hinton}
\affiliation{Hampton University, Hampton, Virginia 23668}
\author{H. Jackson}
\affiliation{Argonne National Laboratory, Argonne, Illinois 60439}
\author{M.K. Jones}
\affiliation{College of William and Mary, Williamsburg, Virginia 23187}
\author{C. Keppel}
\affiliation{Hampton University, Hampton, Virginia 23668}
\affiliation{Physics Division, TJNAF, Newport News, Virginia 23606}
\author{P.H. Kim}
\affiliation{Kyungpook National University, Taegu, Korea}
\author{W. Kim}
\affiliation{Kyungpook National University, Taegu, Korea}
\author{A. Klein}
\affiliation{Old Dominion University, Norfolk, Virginia 23529}
\author{D. Koltenuk}
\affiliation{University of Pennsylvania, Philadelphia, Pennsylvania 19104}
\author{M. Liang}
\affiliation{Physics Division, TJNAF, Newport News, Virginia 23606}
\author{G.J. Lolos}
\affiliation{University of Regina, Regina, Saskatchewan S4S-0A2, Canada}
\author{A. Lung}
\affiliation{Physics Division, TJNAF, Newport News, Virginia 23606}
\author{D.J. Mack}
\affiliation{Physics Division, TJNAF, Newport News, Virginia 23606}
\author{D. McKee}
\affiliation{New Mexico State University, Las Cruces, New Mexico 88003-8001}
\author{D. Meekins}
\affiliation{College of William and Mary, Williamsburg, Virginia 23187}
\author{J. Mitchell}
\affiliation{Physics Division, TJNAF, Newport News, Virginia 23606}
\author{H. Mkrtchyan}
\affiliation{Yerevan Physics Institute, 375036 Yerevan, Armenia}
\author{B. Mueller}
\affiliation{Argonne National Laboratory, Argonne, Illinois 60439}
\author{G. Niculescu}
\affiliation{Hampton University, Hampton, Virginia 23668}
\author{I. Niculescu}
\affiliation{Hampton University, Hampton, Virginia 23668}
\author{D. Pitz}
\affiliation{DAPNIA/SPhN, CEA/Saclay, F-91191 Gif-sur-Yvette, France}
\author{D. Potterveld}
\affiliation{Argonne National Laboratory, Argonne, Illinois 60439}
\author{L.M. Qin}
\affiliation{Old Dominion University, Norfolk, Virginia 23529}
\author{J. Reinhold}
\affiliation{Argonne National Laboratory, Argonne, Illinois 60439}
\author{I.K. Shin}
\affiliation{Kyungpook National University, Taegu, Korea}
\author{S. Stepanyan}
\affiliation{Yerevan Physics Institute, 375036 Yerevan, Armenia}
\author{L.G. Tang}
\affiliation{Hampton University, Hampton, Virginia 23668}
\affiliation{Physics Division, TJNAF, Newport News, Virginia 23606}
\author{R.L.J. van der Meer}
\affiliation{University of Regina, Regina, Saskatchewan S4S-0A2, Canada}
\author{K. Vansyoc}
\affiliation{Old Dominion University, Norfolk, Virginia 23529}
\author{D. Van Westrum}
\affiliation{University of Colorado, Boulder, Colorado 76543}
\author{J. Volmer}
\affiliation{Faculteit Natuur- en Sterrenkunde, Vrije Universiteit, NL-1081 HV 
        Amsterdam, The Netherlands}
\affiliation{NIKHEF, Postbus 41882, NL-1009 DB Amsterdam, The Netherlands}
\author{W. Vulcan}
\affiliation{Physics Division, TJNAF, Newport News, Virginia 23606}
\author{S. Wood}
\affiliation{Physics Division, TJNAF, Newport News, Virginia 23606}
\author{C. Yan}
\affiliation{Physics Division, TJNAF, Newport News, Virginia 23606}
\author{W.-X. Zhao}
\affiliation{M.I.T.--Laboratory for Nuclear Sciences and Department of Physics, 
      Cambridge, Massachusetts 02139}
\author{B. Zihlmann}
\affiliation{University of Virginia, Charlottesville, Virginia 22901}
\affiliation{Physics Division, TJNAF, Newport News, Virginia 23606}
\collaboration{The Jefferson Lab \fpi\ Collaboration}
\noaffiliation

\date{\today}
 
\begin{abstract}
The data analysis for the reaction $^1$H$(e,e'\pi^+)n$, which was used to
determine values for the charged pion form factor \fpi\ for values of
\qsq=0.6-1.6 \gevsq, has been repeated with careful inspection of all
steps and special attention to systematic uncertainties.
Also the method used to extract \fpi\ from the measured longitudinal cross
section was critically reconsidered.  Final values for the separated longitudinal
and transverse cross sections and the extracted values of \fpi\ are presented.
\end{abstract}


\maketitle

\section{Introduction}

Hadron form factors are an important source of information on hadronic
structure.  Of these, the electric form factor, \fpi, of the charged pion plays a
special role.  One of the reasons is that the valence structure of the pion is
relatively simple.  The hard part of the $\pi^+$ form factor can be
calculated within the framework of perturbative QCD (pQCD) as the sum of 
logarithms of \qsq\ multiplied by powers of $1/Q^2$ \cite{lep79}.  As $Q^2
\rightarrow \infty$, only the leading-order term remains
\begin{equation}
F_{\pi}(Q^2\rightarrow\infty ) \rightarrow \frac{16\pi
\alpha_s(Q^2)f_{\pi}^2}{Q^2}
\end{equation}
where $\alpha_s$ is the strong-coupling constant and $f_{\pi}$ the pion decay
constant.
Thus, in contrast to the nucleon case, the asymptotic normalization of the pion 
function is known from the decay of the pion.
The theoretical prediction for \fpi\ at experimentally accessible \qsq\ is less
certain, since the calculation of the soft contributions is difficult and model
dependent.  This is where considerable theoretical effort has
been expended in recent years.  Some
examples include next-to-leading order (NLO) QCD \cite{bak04,melic}, QCD Sum
Rules \cite{radyushkin91,braun}, Constituent Quark Models \cite{hwa01}, and
Bethe-Salpeter Equation \cite{mar00} calculations.  (See Ref. \cite{sterman} 
for a review.)  Some of these approaches
are more model independent than others, but it is fair to say that all benefit
from comparison to high quality \fpi\ data, to delineate the role of hard
versus soft contributions at intermediate \qsq.

The experimental measurement of the pion form factor is quite challenging.  At
low $Q^2$, \fpi\ can be measured in a model independent manner via elastic
scattering of $\pi^+$ from atomic electrons such as at the CERN SPS \cite{ame86}.
Above $Q^2>0.3$ GeV$^2$, one
must determine \fpi\ from pion electroproduction on the proton.
The dependence on \fpi\ enters the cross section via the $t$-channel process,
in which the incident electron scatters from a virtual pion, bringing it
on-shell.  This process
dominates near the pion pole at $t=m_{\pi}^2$.  The physical region for $t$ in
pion electroproduction is negative, so measurements should be performed at the
smallest attainable values of $-t$.  To minimize background contributions, it
is also necessary to separate the longitudinal cross section $\sigma_L$,
via a Rosenbluth L/T(/LT/TT) separation.  The value of $F_{\pi}(Q^2)$ is then
determined by comparing the measured longitudinal cross section at small values
of $-t$, where it is dominated by the $t$-pole term, which contains \fpi,
to the best available model for the reaction $^1$H$(e,e'\pi^+)n$,
adjusting the value of \fpi\ in the latter.

Using the electroproduction technique, the pion form factor was studied for
\qsq\ values from 0.4 to 9.8 \gevsq\ at CEA/Cornell \cite{beb78} and for
\qsq=0.35 and 0.70 \gevsq\ at DESY \cite{ack78,bra79}.  Ref. \cite{bra79}
performed a longitudinal/transverse separation by taking data at two
values of the electron energy. In the experiments done at CEA/Cornell, this was
done in a few cases only, but even then the resulting uncertainties in \sigl\
were so large that the L/T separated data were not used, and \sigl\ was
determined by assuming a certain parametrization for \sigt.  Consequently, the
values of \fpi\ extracted from these data have sizable systematic uncertainties.

More recently, the $^1$H$(e,e'\pi^+)n$ reaction was measured at the Thomas
Jefferson National Accelerator Facility (JLab) in order to study the pion
form factor from \qsq=0.6-1.6 \gevsq.  Because of the excellent properties of
the electron beam and experimental setup at JLab, L/T separated cross sections
were determined with high accuracy.  These data were used to determine the
value of \fpi\ and the results were published in Ref.~\cite{vol01}.  Since
then, the whole analysis chain has been repeated with careful investigation of
all steps, including the contribution of various systematic uncertainties to
the final uncertainty of the separated cross sections.  Furthermore, the method
to determine \fpi\ from the longitudinal cross sections was re-investigated,
leading to a different method to extract \fpi.  In this paper, we report on
these studies and present final results for the longitudinal and transverse
cross sections, as well as the extracted values of \fpi\ from these data.  We
also discuss in detail the extraction of \fpi\ from the measured cross
sections, and the related extraction uncertainties (model dependence).

\section{Experiment and Cross Section Data Analysis}

The cross section for pion electroproduction can be written as
\begin{equation}
\label{eq:sigma1}
\frac{d^3 \sigma}{dE' d\Omega_{e'} d\Omega_\pi} = \Gamma_V J(t)
\frac{d^2 \sigma}{dt d\phi},
\end{equation}
where 
\begin{equation}
\Gamma_v=\frac{\alpha}{2\pi^2} \frac{E^\prime_e}{E_e} \frac{1}{Q^2}
\frac{1}{1-\epsilon} \frac{W^2-M^2}{2 M}
\end{equation}
is the virtual photon flux factor, $\phi$ is the azimuthal 
angle of the outgoing pion with respect to the electron scattering plane,
$t=(p_\pi-q)^2$ is the Mandelstam variable, $J$ is the Jacobian for the
transformation from $d\Omega_\pi$ to $dt d\phi$ and $W$ is the photon-nucleon
invariant mass.

The two-fold differential cross section can be written as
\begin{eqnarray}
\label{eq:sepsig}
2\pi \frac{d^2 \sigma}{dt d\phi} & = & 
   \epsilon \hspace{0.5mm} \frac{d\sigma_{\mathrm{L}}}{dt} +
   \frac{d\sigma_{\mathrm{T}}}{dt} + \sqrt{2\epsilon (\epsilon +1)}
   \hspace{1mm}\frac{d\sigma_{\mathrm{LT}}}{dt}
   \cos{\phi}  \nonumber \\
   & & + \epsilon \hspace{0.5mm}
   \frac{d\sigma_{\mathrm{TT}}}{dt} \hspace{0.5mm} \cos{2 \phi} ,
 \end{eqnarray}
where $\epsilon=\left(1+2\frac{|{\bf
q^2}|}{Q^2}\tan^2\frac{\theta}{2}\right)^{-1}$ 
is the virtual-photon polarization parameter.  The $\sigma_X \equiv
\frac{d\sigma_{\mathrm{X}}}{dt}$ 
cross sections depend on $W$, \qsq\ and $t$.
By using the $\phi$-acceptance of the experiment and taking data for the same
(central) kinematics ($W, Q^2, t$) at two energies, and thus two values of
$\epsilon$, the cross sections \sigl, \sigt, \siglt\ and \sigtt\ can all be
determined. 

Using 2.4-4 GeV electron beams impinging upon a liquid hydrogen
target, data for the $^1$H$(e,e'\pi^+)n$ reaction were taken at a central
value of $W=1.95$ GeV for central $Q^2-$values of 0.6, 0.75, 1.0 and 1.6
\gevsq.  The scattered electron was detected in the Short Orbit Spectrometer
(SOS) and the produced pion in the High Momentum Spectrometer (HMS) of Hall C.

\begin{figure}[t]
\begin{center}
        \includegraphics[width=8.5cm]{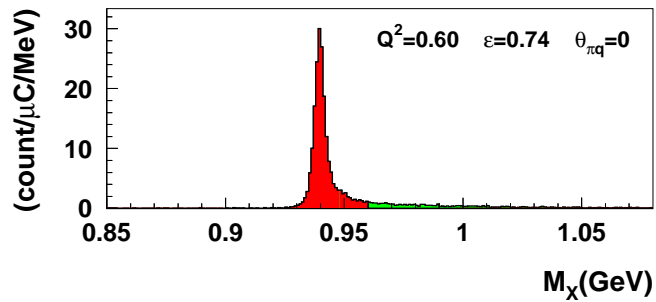}
\end{center}
        \caption{(Color online)
          Representative missing mass distribution.
          Cuts at 0.925 and 0.960 GeV
          were applied to select the recoil neutron final state.
        \label{fig:mx}
        }       
\end{figure}

The data analysis is an updated version of that in Ref.~\cite{vol01}.
First, experimental yields were determined.  Electrons in the SOS were
identified by using the combination of a lead glass calorimeter and gas
\u{C}erenkov detector.  Pion identification in the HMS was accomplished by
requiring no signal in a gas \u{C}erenkov detector and by using time of flight
between two scintillator hodoscope planes.  The momenta of the
scattered electron and the pion at the target vertex were reconstructed from
the wire chamber information of the spectrometers, correcting for energy loss
in the target.  From these, the values of \qsq, $W$, $t$, and the missing mass
were reconstructed.  A cut on the latter of 0.925 to 0.96 GeV was used to
select the neutron exclusive final state, excluding additional pion production
(Fig. \ref{fig:mx}).  Experimental yields as function of \qsq, $W$, $t$ and
$\phi$ were then determined by subtracting random coincidences (varying with
bin but typically 1.2\%) and aluminum target window contributions (typically
0.6\%) and correcting for trigger, tracking and particle-identification
efficiencies, pion absorption, local target-density reduction due to beam
heating, and dead times.  Details of these procedures are similar to those
found in Ref.~\cite{thesis}.

Cross sections were obtained from the yields using a detailed Monte Carlo (MC)
simulation of the experiment, which included the magnets, apertures, detector
geometries, realistic wire chamber resolutions, multiple scattering in all
materials, optical matrix elements to reconstruct the particle momenta at the
target from the information of the wire-chambers of the spectrometers, pion
decay (including misidentification of the decay muon as a pion),
and internal and external radiative processes.

Calibrations with the over-determined $^{1}$H$(e,e'p)$ reaction were
instrumental in various applications. The beam momentum and the spectrometer
central momenta were determined absolutely to better than 0.1\%, while the
incident beam angle and spectrometer central angles were determined with an
absolute accuracy of about 0.5 mrad.  The spectrometer acceptances were checked
by comparison of the data to the MC simulations.  Finally, the overall absolute
cross section normalization was checked. The calculated yields for $e+p$
elastics agreed to better than 2\% with predictions based on a parameterization
of the world data \cite{bos95}.

In the pion production reaction, the experimental acceptances in $W$, \qsq\ and
$t$ are correlated.  By using a realistic cross section model in the MC
simulation, possible errors resulting from averaging the measured yields when
calculating cross sections at average values of $W$, \qsq\ and $t$, can be
minimized.  A phenomenological cross section model was obtained
(see below) by fitting the different cross sections $\sigma_X$ of
Eqn.~(\ref{eq:sepsig}) globally to the data as a function of \qsq\ and $t$ in
the whole range of \qsq.
The dependence of the cross section on $W$ was assumed to follow the
phase space factor $(W^2-M^2_p)^{-2}$, which is supported by previous
data~\cite{bra79}.

The experimental cross sections can then be calculated from the measured and
simulated yields via the relation
\begin{equation}
\label{eq:xs}
\left( \frac{d\sigma(\overline W,{\overline Q}^2,t,\phi)}{dt}
\right)_{\mathrm{exp}} 
= \frac{\langle Y_{\mathrm{exp}}\rangle} {\langle
Y_{\mathrm{MC}}\rangle} 
\left( \frac{d\sigma(\overline W,{\overline Q}^2,t,\phi)}{dt}
\right)_{\mathrm{MC}}.
\end{equation}
This was done for five bins in $t$ at each of the four \qsq-values.  Here,
$\langle Y\rangle$ indicates that the yields were averaged over the $W$ and
\qsq\ acceptance, $\overline W$ and $\overline Q^2$ being the acceptance 
(of high and low \eps\ together) 
weighted average values for that $t$-bin.  By using these
average values, possible errors due to extrapolating the MC model cross
section used to outside the region of the experimental data, is avoided. 

\begin{figure}[t]
\begin{center}
        \includegraphics[width=8.5cm]{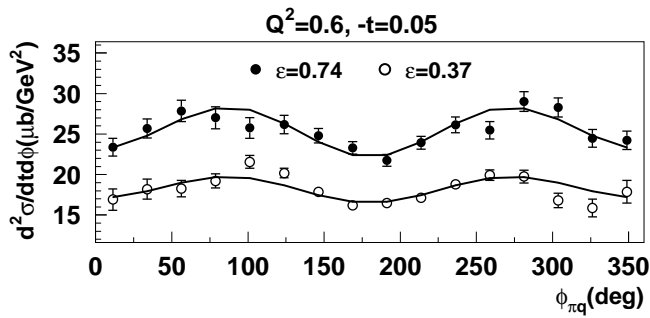}
\end{center}
        \caption{
          Example of the measured cross sections, $\frac{d^2
          \sigma}{dt d\phi}$ as a function of $\phi$ at $Q^2$=0.6 GeV$^2$ for
          two values of $\epsilon$. The curves shown represent the model cross
          section used in the Monte Carlo simulation.
        \label{fig:phi}
        }       
\end{figure}

By combining for every $t$-bin (and for the four values of \qsq) the
$\phi$-dependent cross sections measured at two values of the incoming
electron energy, and thus of \eps, the experimental values
of \sigl, \sigt, \siglt\ and \sigtt\ can be determined by
fitting the $\phi$ and \eps-dependence (Fig. \ref{fig:phi}).
In this fit, the leading order $\sin\theta$
($\sin^2\theta$) of \siglt (\sigtt), where $\theta$ is the angle between the
three-momentum transfer and the direction of the outgoing pion,
was taken into account.\footnote{In the previous
analysis~\cite{vol01}, first \siglt\ and \sigtt\ were
determined by adjusting their values (plus a constant term) until the ratios
$Y_{exp}/Y_{MC}$ were constant as function of $\theta$ and $\phi$.  After that,
\sigl\ and \sigt\ were determined in a Rosenbluth separation.  The present
method is more straightforward and has the advantage that the uncertainties in
the separated cross sections are obtained more directly.}
Those values were then used to improve upon the model cross section used in the
MC simulation.  This whole procedure was iterated until the values of \sigl,
\sigt, \siglt\ and \sigtt\ converged.
The dependence of $\sigma_L$ (and $\sigma_T$) on the MC input model was small
(see below).

\begin{figure}[t]
\begin{center}
        \includegraphics[height=8.5cm,angle=90]{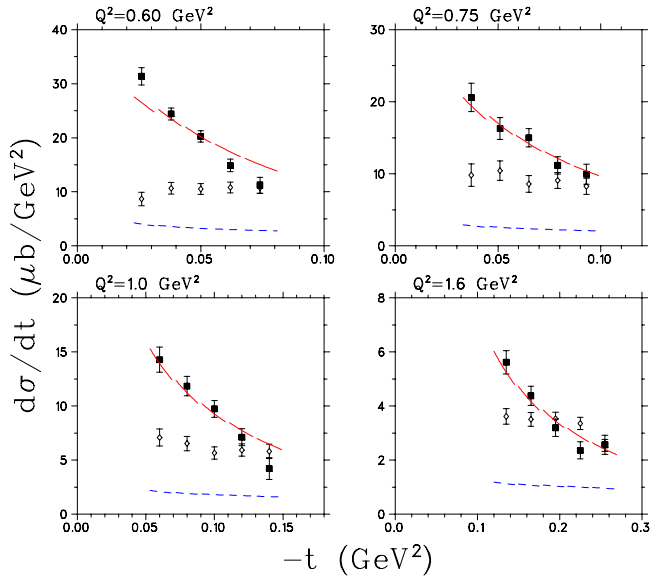}
\end{center}
        \caption{(Color online)
        Separated cross sections \sigl\ [solid] and
        \sigt\ [open].  The error bars represent the combination of
        statistical and $t$ uncorrelated systematic uncertainties.
        In addition, there is an overall systematic uncertainty of about 6\%,
        mainly from the $t$ correlated, \eps\ uncorrelated systematic uncertainty.
        The solid and dashed curves denote VGL model calculations for \sigl\
        and \sigt\ with 
        parameters \Lpi=0.393, 0.373, 0.412, 0.458 \gevsq for \qsq=0.6-1.6
        \gevsq, and with common \Lrho=1.5 \gevsq.
        The discontinuities in the curves result from the
        different average $\overline W$ and ${\overline Q}^2$ of each $t$-bin.
        \label{fig:xsec}
        }       
\end{figure}

The separated cross sections \sigl\ and \sigt\ are shown in Fig.
\ref{fig:xsec}.  They are presented as differential cross
sections $d\sigma /dt$ as a function of $t$, at the
center of the $t$-bin.  The longitudinal cross section exhibits the expected
$t$-pole behavior.  The transverse cross section is mostly flat.

The total uncertainty in the experimental cross sections is a combination of
statistical and systematic uncertainties.  All contributions to the systematic
uncertainty were carefully investigated, also using the results of extensive
single-arm L/T separation experiments and of $^{1}$H$(e,e'p)$ calibration
reactions in Hall C~\cite{chr04}.  The experimental systematic uncertainties
include contributions that, like the statistical uncertainties, are
uncorrelated between the measurements at the two \eps\ values, and others that
are correlated. Most of the uncorrelated ones are common to all $t$ bins, but
there is a small contribution, estimated as 0.7\%, that is also uncorrelated in
$t$.  The $\epsilon$-uncorrelated uncertainties in \sigl\ are inflated by the
factor $1/\Delta\epsilon$ in the L/T separation, where $\Delta\epsilon$ is
the difference (typically 0.3) in the photon polarization between the two
measurements.  The effect on \sigt\ depends on the exact \eps\ values.
The $\epsilon$-uncorrelated systematic uncertainty in the unseparated cross
sections common for all $t$-bins was estimated to be 1.7\%, while the total
correlated uncertainty is 2.8-4.1\%, depending on $t$.  Apart from a
dependence of the separated cross sections on the MC model used, which ranges
from 0.2\% to a maximum of 3\% for one highest $t$-bin, the largest
contributions are: the detection volume (1.5\%), dependence of the extracted
cross sections on the momentum and angle calibration (1\%), target density
(1\%), pion absorption (1.5\%), pion decay (1\%), the simulation of radiative
processes (1.5\%), and detector efficiency corrections (1\%).  The overall
uncertainty is slightly smaller than used in Ref.~\cite{vol01}.

\begin{figure}[t]
\begin{center}
        \includegraphics[width=8.5cm]{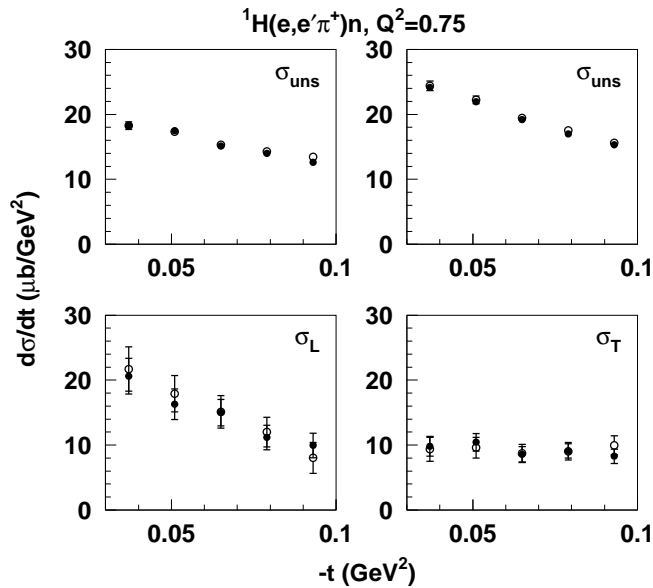}
\end{center}
        \caption{Differential cross section comparison between our earlier
          \qsq=0.75 \gevsq\ results [open circles]
          (Ref. \protect{\cite{vol01}}) and this work [filled circles].  The
          unseparated cross sections ($\sigma_{uns} =
          \epsilon\sigma_L+\sigma_T$) at high and low $\epsilon$ are nearly
          identical, but the differences between the separated \sigl\ and
          \sigt\ are somewhat larger.  The $\sigma_{uns}$ error bars include
          the statistical and epsilon uncorrelated systematics only and are
          in many cases smaller than the plotting symbols.
          Those for $\sigma_{L,T}$ include the contribution of all statistical
          and systematic uncertainties.
        \label{fig:d2comp}
        }       
\end{figure}

\begin{table}
\begin{center}
\begin{tabular}{ccc|c|c}
$\overline{Q^2}$ & $\overline{W}$ & $-t$ & $d\sigma_L/dt$ & $d\sigma_T/dt$\\ 
{\tiny (GeV$^2$)} & {\tiny (GeV)} & {\tiny (GeV$^2$)} & {\tiny ($\mu$b/GeV$^2$)} & {\tiny ($\mu$b/GeV$^2$)}\\ 
\hline
0.526 & 1.983 & 0.026 & 31.360 $\pm$ 1.602, 1.927 &  8.672 $\pm$  1.241\\
0.576 & 1.956 & 0.038 & 24.410 $\pm$ 1.119, 1.774 & 10.660 $\pm$  1.081\\
0.612 & 1.942 & 0.050 & 20.240 $\pm$ 1.044, 1.583 & 10.520 $\pm$  1.000\\
0.631 & 1.934 & 0.062 & 14.870 $\pm$ 1.155, 1.366 & 10.820 $\pm$  0.992\\
0.646 & 1.929 & 0.074 & 11.230 $\pm$ 1.469, 1.210 & 10.770 $\pm$  1.097\\
\hline
0.660 & 1.992 & 0.037 & 20.600 $\pm$ 1.976, 1.895 &  9.812 $\pm$  1.532\\
0.707 & 1.961 & 0.051 & 16.280 $\pm$ 1.509, 1.788 & 10.440 $\pm$  1.344\\
0.753 & 1.943 & 0.065 & 14.990 $\pm$ 1.270, 1.573 &  8.580 $\pm$  1.150\\
0.781 & 1.930 & 0.079 & 11.170 $\pm$ 1.214, 1.416 &  9.084 $\pm$  1.091\\
0.794 & 1.926 & 0.093 & \ 9.949 $\pm$ 1.376, 1.277 &  8.267 $\pm$  1.110\\
\hline
0.877 & 1.999 & 0.060 & 14.280 $\pm$ 1.157, 1.103 &  7.084 $\pm$  0.791\\
0.945 & 1.970 & 0.080 & 11.840 $\pm$ 0.887, 0.978 &  6.526 $\pm$  0.657\\
1.010 & 1.943 & 0.100 & \ 9.732 $\pm$ 0.773, 0.837 &  5.656 $\pm$  0.572\\
1.050 & 1.926 & 0.120 & \ 7.116 $\pm$ 0.789, 0.747 &  5.926 $\pm$  0.570\\
1.067 & 1.921 & 0.140 & \ 4.207 $\pm$ 1.012, 0.612 &  5.802 $\pm$  0.656\\
\hline
1.455 & 2.001 & 0.135 & \ 5.618 $\pm$ 0.431, 0.442 &  3.613 $\pm$  0.294\\
1.532 & 1.975 & 0.165 & \ 4.378 $\pm$ 0.356, 0.390 &  3.507 $\pm$  0.257\\
1.610 & 1.944 & 0.195 & \ 3.191 $\pm$ 0.322, 0.351 &  3.528 $\pm$  0.241\\
1.664 & 1.924 & 0.225 & \ 2.357 $\pm$ 0.313, 0.310 &  3.354 $\pm$  0.228\\
1.702 & 1.911 & 0.255 & \ 2.563 $\pm$ 0.356, 0.268 &  2.542 $\pm$  0.227\\
\end{tabular}
\end{center}
\caption{Separated cross sections \sigl\ and \sigt\ from this work. The two 
listed uncertainties for \sigl\ are the combination of statistical and 
$t$-uncorrelated systematic uncertainties, and the combination of
the $\epsilon$-correlated and uncorrelated uncertainties.  The statistical
and $t$-uncorrelated uncertainties are applied before fitting the VGL model to
the data, while the $\epsilon$-correlated and uncorrelated uncertainties are
applied afterwards.  The listed errors for \sigt\ include all
experimental uncertainties.
\label{tab:dsig}
}
\end{table}

The unseparated cross sections and hence also the values of \sigl\ and \sigt\ of
the present analysis differ from those of our earlier analysis presented in
Ref.~\cite{vol01}.  Compared to that analysis, small adjustments were made in
the values of cuts and efficiencies.  Also, a small mistake was found in
calculating the value of $\theta$, which affects the calculation of the cross
section in Eqn.~\ref{eq:xs}.  Finally, as mentioned, the method to separate the
cross sections was changed.  The cross sections in Table \ref{tab:dsig} are our
final values.  Except for a few cases, the difference with the previous values
is well within the total uncertainty quoted in Ref.~\cite{vol01}.  As an
example, the old and the new cross sections for the case \qsq = 0.75 \gevsq\
are shown in Fig. \ref{fig:d2comp}. It can be seen that the differences in the
extracted unseparated cross sections (top panels) are very small, but the L/T
separation increases them.  On average over the four \qsq\ cases, \sigl\ is 6\%
smaller than in Ref. \cite{vol01} and \sigt\ is 3\% larger.  The largest
differences occur for \qsq=1.0 \gevsq, where \sigl\ is 14\% smaller and \sigt\
is 10\% larger.

\section{Extraction of $\bf F_{\pi}(Q^2)$ from the Data}

It should be clear that the differential cross sections \sigl\ versus $t$ over
some range of \qsq\ and $W$ are the actual observables measured by the
experiment.  The extraction of the pion form factor from these cross sections
can be done in a number of approaches, each with their own merits and
associated uncertainties.

Frazer \cite{frazer} originally proposed that \fpi\ be extracted from
\sigl\ via a kinematic extrapolation to the pion pole, and that this be
done in an analytical manner, \`a la Chew-Low \cite{chew}.  This
extrapolation procedure fails to produce a reliable answer, since different
polynomial fits, each of which are equally likely in the physical region,
differ considerably when continued to $t=m_{\pi}^2$.  Some attempts were made
\cite{kellett} to reduce this uncertainty by providing some theoretical
constraints on the behavior of the pion form factor in the unphysical region,
but none proved adequate.

Bebek et al. \cite{beb78} embraced the use of theoretical input when they used
the Born term model of Berends \cite{berends} to perform a form factor
determination.  Brauel et al. \cite{bra79} similarly used the Born term model
of Gutbrod and Kramer \cite{gut72} to extract $F_{\pi}$.  The presence of the
nucleon and its structure complicates the theoretical model used, and so an
unavoidable implication of this method is that the extraction of the pion form
factor becomes model dependent.

As in Ref.~\cite{vol01,hor06}, the Regge model by Vanderhaeghen, Guidal and Laget
(VGL, Ref.~\cite{van97}) is used here to extract \fpi. 
In this model, the pole-like propagators of Born term models are replaced with
Regge propagators, and so the interaction is effectively described by the
exchange of a family of particles with the same quantum numbers instead of the
exchange of one particle.  
The model was first applied to pion photoproduction.  Most of the
model's free parameters were determined from data on nucleon resonances.
The use of Regge propagators, and the fact that both the $\pi$ ($J=0$) and
the $\rho$ ($J=1$) trajectories are incorporated in the model proved to be
essential to obtain a good description of the $W$- and $t$-dependence
of the photoproduction data for both $\pi^+$ and $\pi^-$ particles.
For electroproduction, the pion form factor and the $\rho \pi \gamma$ form
factor are added as adjustable parameters, parameterized with a monopole form
\begin{equation}
\label{eq:monopole}
F_{\pi}(Q^2) = [1 + Q^2/\Lambda^2_{\pi}]^{-1}.
\end{equation}
The Regge model does a superior job of describing the $t$ dependence of the
differential pion electroproduction cross sections of \cite{bra79,ack78} than
the Born term model.
Over the range of $-t$ covered by this work, \sigl\ is completely determined by
the $\pi$ trajectory, whereas \sigt\ is also sensitive to the $\rho$ exchange
contribution.  The value of \Lrho\ is poorly known, while \Lpi\ is much better
known and in the end is determined by the fitting of the model to the \sigl\
data.

The VGL model for certain choices of \Lpi\ and \Lrho\ is compared to our
data in Figure \ref{fig:xsec}.  The VGL cross sections have 
been evaluated at the same $\overline{W}$ and $\overline{Q^2}$ values as the
data, resulting in the discontinuities shown.
The model strongly underestimates \sigt\ for any value of \Lrho\ used
(variation of \Lrho\ within reasonable values can change \sigt\ by up to 40\%).
Since the JLab data have been taken at relatively low values of $W\approx 1.95$
GeV, this may be due to contributions from resonances, enhancing the strength
in \sigt.  No such terms are included explicitly in the Regge model.  The VGL
model calculation for \sigl\ gives the right magnitude, but the $t$ dependence
of the data is somewhat steeper than that of the calculations. This is most
visible at \qsq=0.6 \gevsq.  As in the case of \sigt, the discrepancy between
the data and VGL is attributed to resonance contributions.  This is supported
by the fact that the discrepancy is strongest at the lowest \qsq\ value, at
higher \qsq\ the resonance form factor reduces such contributions.

Given this discrepancy in shape between the VGL calculations and the \sigl\
data, the questions are: 1) how to determine the value of \fpi\ from the
measured longitudinal cross sections \sigl, and 2) what is the associated
`model uncertainty' in doing so?
The difficulty is that there is no theoretical guidance for the assumed
interfering background.
This applies even if one assumes that the background is due to resonances:
virtually nothing is known about the L/T character of
resonances at $W=1.95$ GeV, let alone their influence on \sigl\ (via
interference with the VGL amplitude). 

\subsection{Summary of our Previous Extraction Method \label{sec:offset}}

In our previous analysis \cite{vol01}, the following procedure was adopted.
When fitting the value of \Lpi\ (and hence \fpi) for the separate $t$-bins
the value of \Lpi\ increases when $-t$ decreases, since the data
are steeper in $|t|$ than the VGL calculations.  The value of \fpi\ extracted
from the lowest $|t|$-bin, which is closest to the pole, was thus taken as
a lower limit.

An upper limit for \fpi\ was obtained by assuming that the background
effectively yields a constant negative contribution to \sigl.  This background
and the value of \Lpi\ were then fitted together, assuming that the background
is constant with $t$.  The fitted contribution of the
background was found to drop strongly as \qsq\ increased from 0.6 to 1.6
\gevsq. Since in \sigt\ this `missing background' (i.e. the difference between
the data and VGL model) decreases, at least for \qsq=0.6 \gevsq, with
decreasing $-t$ (Fig. \ref{fig:xsec}), and assuming that this also holds for
\sigl, these assumptions give an upper estimate for \fpi.  The best estimate
for \fpi\ was then taken as the average of the two results and one half of the
average of the (relative) differences was taken as the `model uncertainty'.

However, the asumptions made in this procedure may be questioned.  Firstly, the
value of \fpi\ extracted from the lowest $|t|$-bin does not have to be a lower
limit, and secondly, the assumption of a negative interfering $t$-independent
cross section in the upper limit calculation requires a special magnitude and
phase for the interfering amplitude with respect to the VGL amplitude.

\subsection{Another Form Factor Extraction Method \label{sec:ampl}}

Since the publication of those results, we have looked at the discrepancy
between the $t$-dependence predicted by the VGL model and the data in more
detail by assuming, besides the VGL amplitude, a $t$-independent interfering
background {\it amplitude}, and fitting the latter together with the
value of \Lpi.  The fitting uncertainty in \Lpi\ varies between 5\% and 18\%,
while the magnitude and phase of the fit background amplitude are very poorly
constrained (uncertainties in the hundreds of percent).

Although the fitting uncertainties are very large, the results of this exercise
suggest an interfering amplitude whose magnitude decreases monotonically with
increasing \qsq, but whose phase with respect to the VGL amplitude does not
necessarily result in a net negative cross section contribution to \sigl, as
has been assumed in the previous analysis.
However, here also a special assumption was used, i.e. an interfering
amplitude with a magnitude and phase that do not depend on $t$.
Thus, determining \fpi\ in this way is not a viable method, either.

\begin{figure}[h]
\begin{center}
        \includegraphics[height=8.5cm,angle=90]{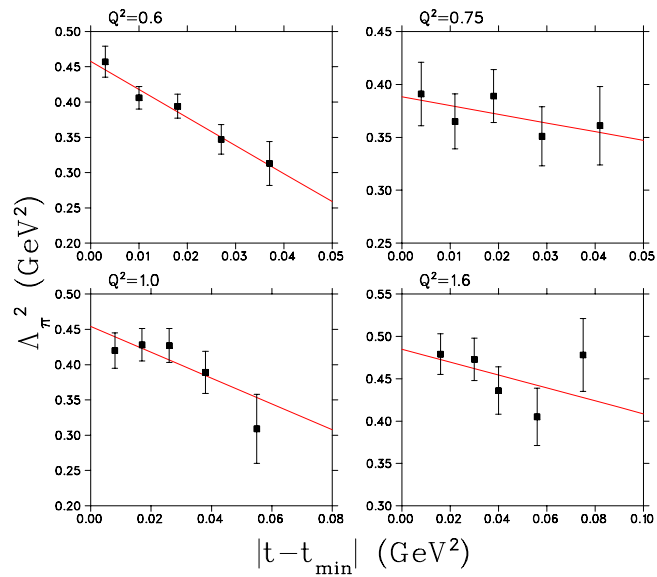}
\end{center}
        \caption{(Color online)
        Values of \Lpi\ determined from the fit of the VGL model to each
        $t$-bin, and linear fit to same.  
        The error bars reflect the
        statistical and $t$-uncorrelated systematic uncertainties.  The additional
        overall systematic uncertainties, which are
        applied after the fit, are not shown.  
        \label{fig:lpi}
        }       
\end{figure}

\subsection{Adopted Form Factor Extraction Method \label{sec:tmin}}

Given that no information is available on the background, we are forced to make
some assumptions in extracting \fpi\ from these data.  Our guiding principle
is to minimize these assumptions to the greatest extent possible.  The
form factor extraction method that we have adopted relies on the single
assumption, that the contribution of the background is smallest at the kinematic
endpoint $t_{min}$.

\begin{table}
\begin{center}
\begin{tabular}{cc|c|c}
$Q^2$ & W & \Lpi\ & $F_{\pi}$ \\ 
(\gevsq) & (GeV) & (\gevsq) & \\ \hline
0.60   & 1.95  & $0.458\pm 0.031^{+0.255}_{-0.068}$ & $0.433\pm 0.017^{+0.137}_{-0.036}$ \\
0.75   & 1.95  & $0.388\pm 0.038^{+0.135}_{-0.053}$ & $0.341\pm 0.022^{+0.078}_{-0.031}$ \\
1.00   & 1.95  & $0.454\pm 0.034^{+0.075}_{-0.040}$ & $0.312\pm 0.016^{+0.035}_{-0.019}$ \\
1.60   & 1.95  & $0.485\pm 0.038^{+0.035}_{-0.027}$ & $0.233\pm 0.014^{+0.013}_{-0.010}$ \\ 
\hline
0.70   & 2.19  & $0.627\pm 0.058^{+0.096}_{-0.085}$ & $0.473\pm 0.023^{+0.038}_{-0.034}$ \\
\end{tabular}
\end{center}
\caption{\Lpi\ and \fpi\ values from this work, and the reanalyzed data from
   Ref. \protect{\cite{bra79}} using the same method.
   The first error includes all experimental and analysis uncertainties,
   and the second error is the `model uncertainty' as described in the text.
\label{tab:fpi}
}
\end{table}

Our best estimate for \fpi\ is thus determined in the following manner.
Using the value of \Lpi\ as a free parameter, the VGL model was fitted to each
$t$-bin separately, yielding $\Lambda^2_\pi(\overline{Q^2},\overline{W},t)$
values as shown in Fig. \ref{fig:lpi}.
The values of \Lpi\ tend to decrease as $-t$ increases, presumably
because of an interfering background not included in the model.
Since the pole cross section containing \fpi\ increases strongly with
decreasing $-t$, and the background presumably remains approximately constant,
as suggested by the difference between data and VGL calculations for \sigt, we
assume that the effect of this background will be smallest at the smallest
value of $|t|$ allowed by the experimental kinematics, $|t_{min}|$. Thus, an
extrapolation of \Lpi\ to this physical limit is used to obtain our best
estimate of \fpi.  The value of \Lpi\ at $t_{min}$ is obtained by a linear fit
to the data in Fig. \ref{fig:lpi}.
The resulting \Lpi\ and \fpi\ values are listed in Table
\ref{tab:fpi}.  The first uncertainty given represents both the experimental 
and the linear fit extrapolation uncertainties.

The \fpi\ values listed in Table \ref{tab:fpi} correspond to the true values
within the context of the VGL model if, and only if, the background vanishes
at $|t-t_{min}|=0$.  Because of the uncertainty inherent in this assumption,
we also estimate a `model uncertainty' to account for the possible influence
of the missing ingredient in the VGL model (background) at $|t-t_{min}|=0$.  
Lacking a model for the background, we can only try to make a fair estimate of
this uncertainty.  This was done by looking at the variation in the fitted
values of \Lpi\ when using two different assumptions for the background.  We
used the two cases considered earlier in this paper when trying to determine
\fpi.  The first case assumes the $t$-independent negative background in
addition to the VGL model used in Sec. \ref{sec:offset}.  The second case
assumes the interfering background amplitude with a $t$-independent magnitude
and phase discussed in Sec. \ref{sec:ampl}.  However, here they are not used to
determine \fpi, but only to estimate the model uncertainty in our best value of
\fpi\ determined above.

The estimated model uncertainty is determined from the spread of the \Lpi\ values
at each \qsq\ given by these two methods.  Each effectively represents a
different background interference with the VGL model.  To keep the
number of degrees of freedom the same in all cases, the background was fixed to
the minimum $\chi^2$ value determined in each of the above two studies, and
\Lpi\ and its uncertainty was then determined in a one-parameter fit of the VGL
model plus background to the \sigl\ data.  Since there is a strong
statistical overlap between the two fits, the statistical plus random
uncertainties of the data were quadratically removed from the \Lpi\
uncertainties.  The model uncertainty at each \qsq\ is then assigned to be the
range plus fitting uncertainty given by the two methods, relative to the value
of \Lpi\ at $t_{min}$.  The resulting (asymmetric) uncertainties are listed as
the second uncertainty in Table \ref{tab:fpi}.

The model uncertainty in the \fpi\ value drops from about 20\% at \qsq=0.6
\gevsq\ to about 5\% at 1.6 \gevsq.  This is
consistent with the fact that the discrepancy with the $t$-dependence of the
VGL calculation is smaller for the larger values of \qsq. It is
also at least compatible with the idea that resonance contributions, which
presumably have a form factor that drops fast with \qsq, are responsible.  
The corresponding model uncertainty quoted in
Ref.~\cite{vol01} was approximately 8.5\% at all \qsq, but that was based on a
more restrictive assumption on the background (in essence case 1).

\section{Results and Discussion}

\begin{figure}[h!]
\begin{center}
        \includegraphics[height=8.5cm,angle=90]{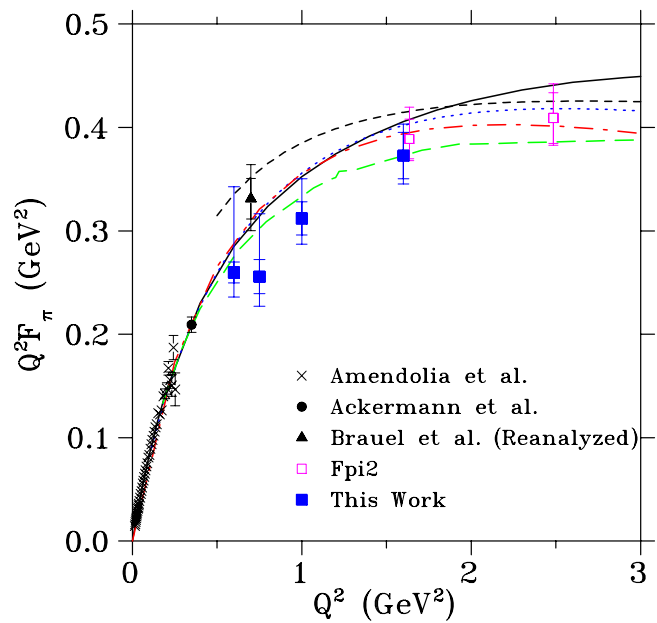}
        \caption{(Color online)
        $Q^2F_{\pi}$ data from this work, compared to previously published data.
        The Brauel et al. \protect{\cite{bra79}} point has been reanalyzed
        using the \fpi\ extraction method of this work.
        The outer error bars for this
        work and the reanalyzed Brauel et al. data include all experimental
        and model uncertainties, added in quadrature, while the inner error
        bars reflect the experimental uncertainties only.
        Also shown are the 
        light front quark model \protect{\cite{hwa01}} (dash-dot), 
        Dyson-Schwinger \protect{\cite{mar00}} (solid), 
        QCD sum-rule \protect{\cite{nes82}} (dot), 
        dispersion relation \protect{\cite{ges00}} (long-dash), 
        and quark-hadron duality \protect{\cite{mel03}} (short-dash),
        calculations.
        \label{fig:q2fpi}
        }       
\end{center}
\end{figure}

Because of the arguments given above, the values presented in
Table~\ref{tab:fpi} and Fig.~\ref{fig:q2fpi} are our final estimate of
\fpi\ from these data using this model.  However, we stress that the primordial
results of our experiment are the \sigl\ cross sections.  When improved models
for the $^1$H$(e,e'\pi^+)n$ reaction become available, other (better) values of
\fpi\ may be extracted from the same cross sections.

The present values for \fpi\ are between 7\% and 16\% smaller than our previously
published values~\cite{vol01}, which is about the combined experimental and
model uncertainty.  The largest difference is at $Q^2=0.75$ \gevsq.  On
average, one quarter of the difference is because the final values of \sigl\
are smaller than those of Ref.~\cite{vol01} (see Fig. \ref{fig:d2comp} for a
representative comparison), and the remaining three quarters are due to the
\fpi\ extraction method, the present method being closer to the method used in
Ref.~\cite{vol01} to obtain the lower limit.

Analyses of other data at higher $W$ indicate that the discrepancy with the
$t$-dependence of the VGL calculation is much smaller at higher values of $W$.
The data from Brauel et al.~\cite{bra79}, taken at \qsq=0.70 \gevsq\ and a
value of $W$=2.19 GeV, were reanalyzed using the \fpi\ extraction method
presented here.  The result is 0.4\% higher than that obtained using the \fpi\
extraction method of Ref. \cite{vol01}.  This indicates that our \fpi\
extraction methods are robust when the background contribution is small, as
appears to be the case at this higher value of $W$.

The data from our second experiment \cite{hor06} at $W$=2.22 GeV and \qsq=1.6,
2.45 \gevsq\ are also shown in Fig. \ref{fig:q2fpi}.  There, the VGL model
adequately describes the $t$-dependence of the \sigl\ data, again indicating
that the background contributions for \sigl\ are smaller at higher $W$,
even though the model strongly underpredicts the magnitude of \sigt. 
In that case, the VGL model was fit to the full $t$-range of the \sigl\ data
with only small fitting uncertainties.
It is seen in the figure that the revised \qsq=1.6 \gevsq\ result
from this work agrees well with that from our second experiment,
taken at higher $W$ and 30\% closer to the $\pi^+$ pole.  The excellent
agreement between these two results, despite their significantly different
$t_{min}$ values, indicates that the uncertainties due to the $\pi^+$
electroproduction reaction mechanism seem to be under control, at least in this
\qsq-range.

Fig.~\ref{fig:q2fpi} compares our final data from this work and from our second
experiment \cite{hor06} to several QCD-based calculations.  The combined
data sets are consistent with a variety of models.  Up to \qsq=1.5 \gevsq, the
Dyson-Schwinger calculation of Ref. \cite{mar00}, the light front quark model
calculation of Ref.~\cite{hwa01}, and the QCD sum-rule calculation of
Ref.~\cite{nes82} are nearly identical, and are all very close to the monopole
form factor constrained by the measured pion charge radius~\cite{ame86}.  Such
a form factor reflects non-perturbative physics.
Our revised data are below the monopole curve.  A significant deviation would
indicate the increased role of perturbative components at moderate \qsq, which
provide in that region a value of $Q^2F_{\pi}\approx$0.15-0.20 only
\cite{bak04}.  The dispersion relation calculation of Geshkenbein et
al. \cite{ges00} is closer to our results in the \qsq=0.6-1.6 \gevsq\ region,
while still describing the low \qsq\ data used for determining the pion
charge radius.
The quark-hadron duality calculation by Melnitchouk \cite{mel03} is not
expected to be valid below \qsq=2.0 \gevsq.  This is reflected in its
significant deviation from the monopole curve at low \qsq.  To better
distinguish between these different models, it is clear that especially higher
\qsq\ data, as well as more data at higher values of $W$ in the \qsq=0.6-1.6
\gevsq\ region, are needed.  Plans are underway to address both of these at
JLab.

\section{Summary and Conclusions}

To summarize, the data analysis for our $^1$H$(e,e'\pi^+)n$ experiment at
\qsq=0.6-1.6 \gevsq, centered at $W=1.95$ GeV, has been repeated with careful
inspection of all steps.  The final unseparated cross sections presented here
are in most cases consistent with our previous analysis within
experimental uncertainties.  After the magnifying effect of the L/T
separation, the resulting \sigt\ values are slightly larger than
before, and the \sigl\ values are correspondingly smaller.
The experimental systematic uncertainties were critically reviewed, and are
slightly smaller compared to the previous analysis.

As before, we use a fit of the Regge model of Ref. \cite{van97} to our \sigl\
data to extract \fpi.  The data display a steeper $t$-dependence than the
model, which we attribute to the presence of longitudinal background
contributions not included in the model.  After revisiting our prior
assumptions used to extract \fpi\ from \sigl\ with the model, we conclude that
some of our prior analysis assumptions were unwarranted.  Therefore, we employ a
revised \fpi\ extraction method which relies only on the assumption that the
background contributions are minimal at $t_{min}$.  
The resulting values are our best estimate of \fpi\ from these data with this
model, and are between 8\%
and 16\% smaller than before, primarily due to the different extraction method.
The Brauel et al. \cite{bra79} data at similar \qsq\ but higher $W$
are robust against our fitting assumptions, consistent with our
expectation that a longitudinal background contribution not included in the
Regge model is the cause of the discrepancy.

The new analysis, in addition to providing our final \fpi\ results
for the \qsq=0.6-1.6 \gevsq\ range, gives an indication of the contribution of
the analysis assumptions to
the \fpi\ determination.  A detailed analysis yields model uncertainties that
decrease with increasing \qsq\ and $W$.  They are consistent with
the differences in the values of \fpi\ determined using the previous and
present extraction methods.  They indicate that, given the present
electroproduction model,
the uncertainty in the dtermination of \fpi\ in this $W$ and \qsq\ range is of
the order of 10\%.

The revised data indicate that for \qsq$>0.5$ \gevsq, \fpi\ starts to fall
below the monopole curve that describes the low \qsq\ elastic scattering data.
These results are consistent with those of our second, more precise experiment
at higher \qsq and $W$ \cite{hor06}.  The two sets of data at $Q^2=1.6$ \gevsq\
are taken with significantly different $t_{min}$, and so if the various
form factor extraction issues were not being handled well by the VGL model, a
significant discrepancy would have been expected to result.  Their good
agreement lends further credibility to the analysis presented here.  
It will be useful to acquire additional electroproduction data in the
$0.5<Q^2<1.5$ \gevsq\ range at higher $W$ in order to be able to extract more
precise values of \fpi\ without the difficulties encountered here.

\section{Acknowledgments}

The authors would like to thank Drs. Guidal, Laget and Vanderhaeghen for
stimulating discussions and for modifying their computer program for our needs.
This work is supported by DOE and NSF (USA), FOM (Netherlands), NSERC (Canada),
KOSEF (South Korea), and NATO.

\end{document}